# Particle production in proton-proton collisions

M. T. Ghoneim[A], F. H. Sawy[B] and M. T. Hussein[A]

[A] Physics Department, Faculty of Science, Cairo University, Giza – Egypt
[B] High Energy Physics (HEP) lab, Physics Department, Faculty of Science, Cairo University, Giza – Egypt

ghoneim@sci.cu.edu.eg, fatma.helal.sawy@cern.ch and tarek@Sci.cu.edu.eg

**ABSTRACT**

Proton-proton collision is a simple system to investigate nuclear matter and it is considered to be a guide for more sophisticated processes in the proton-nucleus and the nucleus-nucleus collisions. In this article, we present a phenomological study of how the mechanism of particle production in pp interaction changes over a wide range of interaction energy. This study is done on data of charged particle produced in pp experiments at different values of energy. Some of these data give the created particles classified as hadrons, baryons and mesons, which help us compare between production of different particles. This might probe some changes in the state of nuclear matter and identify the mechanism of interaction.

*Keywords*:  string - gluon fragmentation –jet production.

# 1-INTRODUCTION

Particle production, the concept of mass-energy relationship, has been always one of the selected topics to investigate in high energy nuclear reactions over several decades. This type of research has probably started by the time people wanted to accelerate particles up to relativistic speeds and to smash them into other particles and see what may turn out. Passing over the techniques of acceleration and particle detection, people observed that in proton-proton (pp) collisions at relativistic energy, more particles came out than those went in. The extra came out particles were, principally, created pions and/or heavier particles at higher interaction energies. Our main task in this work is to follow up, qualitatively, the variation of the average multiplicity and distribution of created particles with the reaction energy by making use of the available experimental data and their corresponding theoretical aspects over a wide range of interaction energy; from few GeV up to several TeV.

## 2-PARTICLE CREATION BETWEEN THEORY AND EXPERIMENT:

The physical nature of proton-proton interactions varies with energy as a result of the decrease of the coupling constant with energy. This fact could lie behind the inability to find a unique theory to describe particle creation mechanism in p-p interactions over the whole available energy range. Such creation goes through phenomenological models in low energy region to perturbative quantum chromo-dynamics (PQCD) in high energy one (Daniel et al., 2003, Adler et al. 2003, Regge 1959 and Gribov 1967).

In view of one of the most useful models in this subject; the "String Model", two protons, each containing three valence quarks, interact by color field exchange which resembles a string. The collision energy divides each proton into its constituent quarks to form strings between the quarks (diquarks) from the projectile and the target. The string forces between the color charges cause periodical oscillations of the system and the color field may materialize at a point of the string. As time develops, the string breaks randomly into smaller pieces carrying smaller fraction of the initial energy. The primary string is formed among the originally interacting quarks while a secondary string is generated in the field of sea quarks of the primary string and so on for higher order strings. Assuming z to be a fraction of energy carried by a string, it continues producing particles in a ranking order governed by a fragmentation function, $f(z)$, until the residual energy becomes less than the threshold of production. Some string dynamics (M.T.Hussein et al. 1995) that consider higher orders, assume that a jet is formed by an original quark $q_0$, initially with energy $w_0$, where at the first vertex of fragmentation, the string breaks forming a $q\bar{q}$ with energy $zw_0$ leaving the string with energy $w_1 = (1-z)w_0$, where $0 < z < 1$ with the distribution $f(z)$. Further fragmentation of the string would produce $q\bar{q}$ pairs forming particles whose $i^{th}$ rank order having energy $zw_{i-1}$. The length of the string would depend upon the quark energy $w_0$ and $f(z)$. One would expect that the longer the string, the larger the number of vertices and the higher the multiplicity of the formed mesons. If $z_{av}$ is the average value of the parameter z, then the residual energy of the string after the $n^{th}$ rank would be:

$$w_n = (1 - z_{av})^{\bar{n}} w_0 \qquad (1)$$

,where

$$\bar{n} = \ln(w_0/w_n)\beta \qquad (2)$$

With:
$$\beta = -1/\ln(1 - z_{av}) \qquad (3)$$

, $\bar{n}$, is the average multiplicity of the produced particles. When the mass of a string piece gets small enough, it is identified as a hadron and the breaking stops within that piece meaning that the whole system eventually evolves into hadrons (hadronization process). Figure (1) presents the dependence of the average multiplicity on the the center of mass energy, s, over the whole range of data. Crosses in this figure are experimental data (Badawy et al. 2008, Minami 1973, http.//pdg.lbl.gov, Biyajima et al. 2001, Wolschin 2001) while the blue curve is their polynomial fit:

$$\bar{n} = a (\ln s)^2 + b \ln s + c \qquad (4)$$

The values of the fitting parameters; a, b and c are: 0.29, -1.57 and 5.61.

One may start with the "softly" created particles, where the average multiplicity, in the lower energy portion of figure (1), shows a linear logarithmic relationship with the energy the fit of which is given by the straight line with fitting parameters 0, 1.0 and -3.66 in equation (4), respectively. This linearity, that does not describe the data in the higher energy region (as its extension shows) is supported by the Fermi scaling, as well as by most of theoretical models like the "multi-peripheral model" and the "quark-parton model" in asymptotic soft regions (Cerny et al. 1977, and Fiete et al. 2010). The parameterization works reasonably and this linearity fits well up to several tens of GeV. The multiplicity distribution of created particles by soft events follows a Poisson distribution;

$$P(n) = \frac{\bar{n}^n}{n!} e^{-\bar{n}} \qquad (5)$$

which means that every single final-state particle is created and emitted independently and viewed as a black body radiation (Niccol`o et al., 1998). The KNO scaling of the multiplicity distribution (within this energy range) for non-single diffractive NSD events in full phase space supports a single creation mechanism (Koba et al. 1972).

Above the last stated energy range, the multiplicity distribution shows a deviation from the Poisson shape, predicting some kind of correlations between the created particles and a sign of variation in the creation mechanism. These correlations were proposed by the "Clan Model" (Fiete et al. 2010), that assumes the ability of a particle to emit additional particles, as cascading by decay and fragmentation. The model considers that the created particles stream as clans (clusters) where the ancestors production, and thus the clans, are governed by a Poisson distribution. A clan contains all particles that stem from the same ancestor, where the ancestors themselves are produced independently. The interpretation of "Clan Model" was based on the success of the Negative Binomial Distribution NBD of the particle multiplicity:

$$P(n; \bar{n}; k) = \binom{n+k-1}{n} \left(\frac{\bar{n}/k}{1+\bar{n}/k}\right)^n \frac{1}{(1+\bar{n}/k)^k} \qquad (6)$$

, which describes the multiplicity distributions up to √s = 540 GeV, announced by UA5 (Giovannini et al. 1986). The NBD is defined by two parameters n̄ and k. n̄ is the average multiplicity as mentioned above and the parameter k is lelated to the dispersion D as:

$$\frac{D^2}{(\bar{n})^2} = \frac{1}{\bar{n}} + \frac{1}{k} \qquad (7)$$

For (1/k) → 0, the NBD reduces to the Poisson distribution, and for k = 1 it is the geometric distribution. It was found that (1/k) increases almost linearly with ln(s) whereas KNO scaling corresponds to a constant, energy-independent.

The same meaning could be found in the string model that proposed multi-order fragmentation for higher energy. The interpretation had got more support by assuming new type of events, called "semi-hard" in addition to "soft" ones to produce these bundles. Experimentally, semi-hard events are responsible for a "mini-jet" production (Niccol`o et al., 1998). A "mini-jet" is defined as a group of particles having a total transverse momentum larger than 5 GeV/c (Tran et al. 1988, Giovannini and Ugoccioni 1999). As energy goes higher than 540 GeV; "semi-hard" events start to show significant contribution in collision and many models were modified by adding new terms. The Dual Parton Model (DPM) claims that "mini-jets" are generated from at least four chains, two of them come from a contribution of valence quarks and the other two are generated from sea quarks through semi-hard interaction. As energy goes higher, sea quark contribution goes bigger. Since sea quarks carry only a small fraction of the momentum of the incident hadrons, the chains are concentrated in the central rapidity region. Thus, these may explain the rise of the central particle density. Consequently the KNO scaling has been violated in this region. This violation is traced to short range correlations of particles in the strings and interplay between the double-pomeron processes (Niccol`o et al., 1998, Giovannini and Ugoccioni 1999). The superposition of the two types of interaction affects the multiplicity distribution and therefore potentially explains the deviation from the scaling found at lower energies. UA5 has been successfully fitted the multiplicity distributions of created particles as a superposition of two independent NBDs, at √s = 900 GeV up to 1800 GeV, and this is supported by a two-component model (Niccol`o et al., 1998 and Giovannini and Ugoccioni 1999). The quadratic term was then found to be a better description of the data as the interaction energy goes above the limits of pure soft events stated before. This term reflects the contribution of semi-hard and gluon-bremsstrahlung process, that starts to manifest it self as the interaction energy gets into the TeV region (Giovannini and Ugoccioni 1999). These radiation are known as Initial and Final State Radiations that emit gluon before and after real collision has occurred, respectively. These radiation (gluon) can materialize to produce hadrons and the increase in collision energy, increases the contribution of this radiation in particle production.

As energy goes higher, the strong coupling constant becomes smaller and smaller, slipping into asymptotic freedom. Collisions of pp at this higher energy can be viewed as quark-quark collisions which can be mathematically described by perturbative quantum chromo-dynamics (PQCD). These high energy quark collisions generate new type of events called "hard" events in addition to "soft" and "semi-hard" ones. Hard quark interactions develop via short-distance over a very short time scale and the subsequent fragmentation produces a cone of hadronic final states that originate from the same quarks. This cone of hadrons is called a "jet", representing an independent fireball for hadron creation, and the properties of the jet depend only on the initial quark. The proposal of DPM is forming more than four chains by multiple pomeron exchanges, Figure (2), and increase the height of the central plateau with energy. Therefore, the multi-chain contribution becomes increasingly important and the average number of chains increases with energy. The growth of multiplicity at higher energy can be understood by assuming that; as the interaction energy goes higher; gluon jets grow with higher multiplicity than the quark jets and both compete with each other to, eventually, produce particles through their multifragmentations. QCD predicts that gluon initiates jets to have higher average particle multiplicity compared to quark initiated ones (CMC collaboration 2012). This is supported fairly by the red curve, in figure (2), that results from the subtraction of the staright line (which represents the softly produced multiplicity) from the total curve, to obtain the multiplicity relation that belongs to the semi-hard and hard mechanism events. One might add, at this point, that production of particles in the nuclear reaction is one of the macroscopic parameters to probe what is going on inside the reacting systems during the reaction and the development of the processes that take place in these systems as the interaction energy goes higher. The red cuve in figure (1) indicates some phase change of the nuclear matter and consequently of the mechanism by which different types of particles are created. This curve also shows the limit ($\sqrt{s}$ = 53 GeV) at which such a change started to occur. Some workers (STAR Collaboration, 2014) has defined the hard event as that having at least one jet cluster while the soft one as that having no clusters at all they also add that the relation between $< P_T >$ and the multiplicity, in soft events has a weak dependence of the collision energy from the RHIC to the Tevatron and the properties of the final states are determined only by the number of the charged particles, while hard events have much stronger dependence. A recent study has described the multiplicity – energy relationship by a linear, quadratic and cubic terms in a polynomial. This study interpreted the second and third orders in their fitting polynomial as coming from double and triple quark interactions, respectively, rather than from only one (Ashwini et al. 2013, Alexopoulos et al. 1998 and Walke et al. 2004).

In the above part of this work, so far, we have considered creation of particles without distinguishing between their entities. Let us shed some light on different types of created particles taken separately from different experimental data that have reasonable statistics and rational consistency (Antinucci et al.1973, Rushbrooke et al.1982, Samset et al.2006, Engel 2008, Ansorge et al. 1989, Anticic et al. 2010 and Becattini et al 1997). The dependence of the mean multiplicity of created pions, keons and lambdas, $\bar{n}_\pi$, $\bar{n}_k$, $\bar{n}_\Lambda$; respectively, on ln(s), is displayed in figure (3) and seems to follow a similar polynomial to that in equation (4) with fitting paprameters given in table (1). Figure (3) shows that below certain values of energy, the data do not show any production of pions, keons and lambdas and each type of these particles starts to appear as energy goes higher past certain values. This is easily understood if one takes into account the threshold of production of each type of these particles. One may notice also that allover the considered energy range, production of pions is more dominant than keons and lambdas. This comes from the fact that creation of pions is more probable than creation of heavier particles i.e. the probability of u and d quarks is higher than s ones, which are responsible for the formation of heavier particles, in spite of the presence of energy that covers the thresholds of their creation. This fact, in turn, is probably a consequence of the color field nature that loses its energy bit by bit as excitations (kinks) of soft gluons rather than by hard single gluon radiation (Niccol`o et al., 1998, Tai An et al.and Sa Ben-Hao 1998, Greiner et al. 1994 and Ellis et al. 1996). Besides, pions are much more stable than heavier particles which decay quickly, also, to pions.

**Table (1): fitting paprameters of Eq (4), with the experimental data of the created particles, namely, pions, keons and lambdas.**

| Created particle | Fitting parameters | | | |
|---|---|---|---|---|
| | a | b | c | d |
| Pions (red curve) | 0.02 | -0.35 | 3.42 | -6.18 |
| Keons (blue curve) | 0 | -0.09 | 0.72 | -1.53 |
| Lambdas (black line) | 0 | 0 | -0.03 | 0.17 |

The relative probability of production of keons and lambdas to that of pions ($R_{\pi/k} = \bar{n}_\pi/\bar{n}_k$ and $R_{\pi/\Lambda} = \bar{n}_\pi/\bar{n}_\Lambda$) over the available energy range are displayed in figure (4). $R_{\pi/\Lambda}$ shows remakable higher values than $R_{\pi/k}$ over all the available energy range, in agreement with figure (3) and one may notice also that $R_{\pi/k}$ decreases with energy, while $R_{\pi/\Lambda}$ is almost constant. This might reflect that formation of different particles results from some changes of the nuclear matter state, which in turn, imposes different mechanisms for creating different particles e.g. the increased enhancement of gluon interactions around the TeV order of interaction energy (LHCb collaboration 2012).

What presented here in this work, was just a general survey of qualitative descriptions of particle creation in the pp collision as a preliminary work for a next coming study that would go deeper in this open uncomplete subject, including more mathematical and computational details.

## **CONCLUSION**

Along the studied energy range, the experimental data of created particles in proton proton interaction show different mechanisms that could be divided into three parts: soft, semi-hard and hard components. The soft component, which means that the final-state particles are created and emitted uncorrelated, seems to exist allover the considered range of energy. It is probably the only mechanism in particle creation up to several tens of GeVs. This nature changes as energy passes through several hundreds of GeVs region, where particles have the ability to emit additional particles by decay and cascade production, thus adding a semi-hard component to the soft one, in the creation process. This trend seems to indicate the existence of new phases in the interacting systems, presumably that the average number of created particles represents a macroscopic parameter eligible to probe the state of the system and its variation with interaction energy. On approaching the borders of the TeV region, a hard component takes a leading role in creation in addition to the other components, as a result of the growth of the gluon interactions. Heavier particle production (like keons and lambdas) show remarkable lower production rate than pion. This might be due to the small bits of energy losses of the color field as excitation of soft gluons rather than by hard single gluon radiation. Beyond these limits, the results show that production of heavier particles than pions flourishes as the interaction energy grows up, where gluons jet, fragmentations and long reformations at enough energies furnish more suitable conditions for heavy quarks creation than quark jets do.

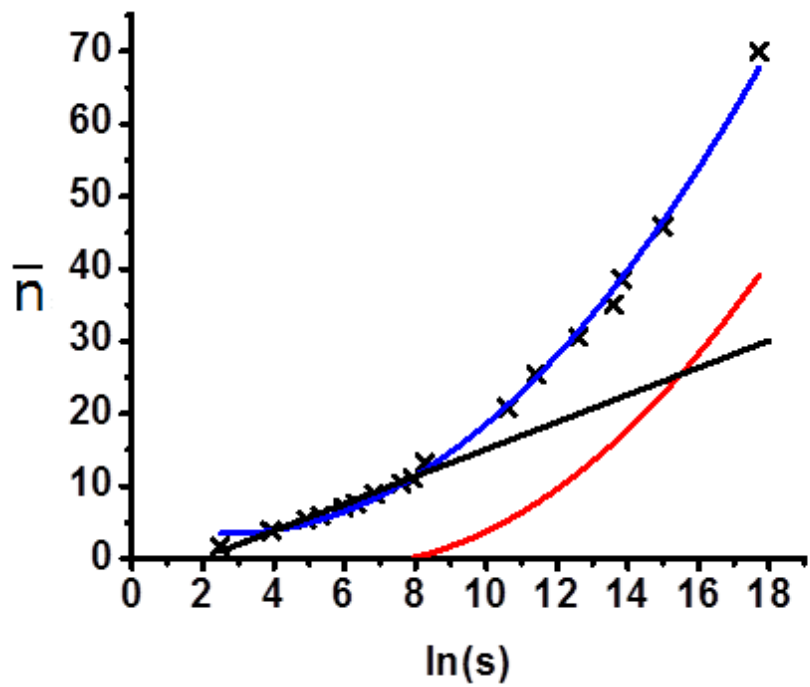

**Fig (1)**

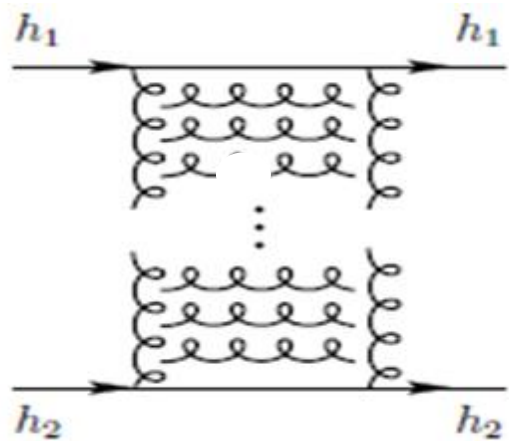

**Fig.2**

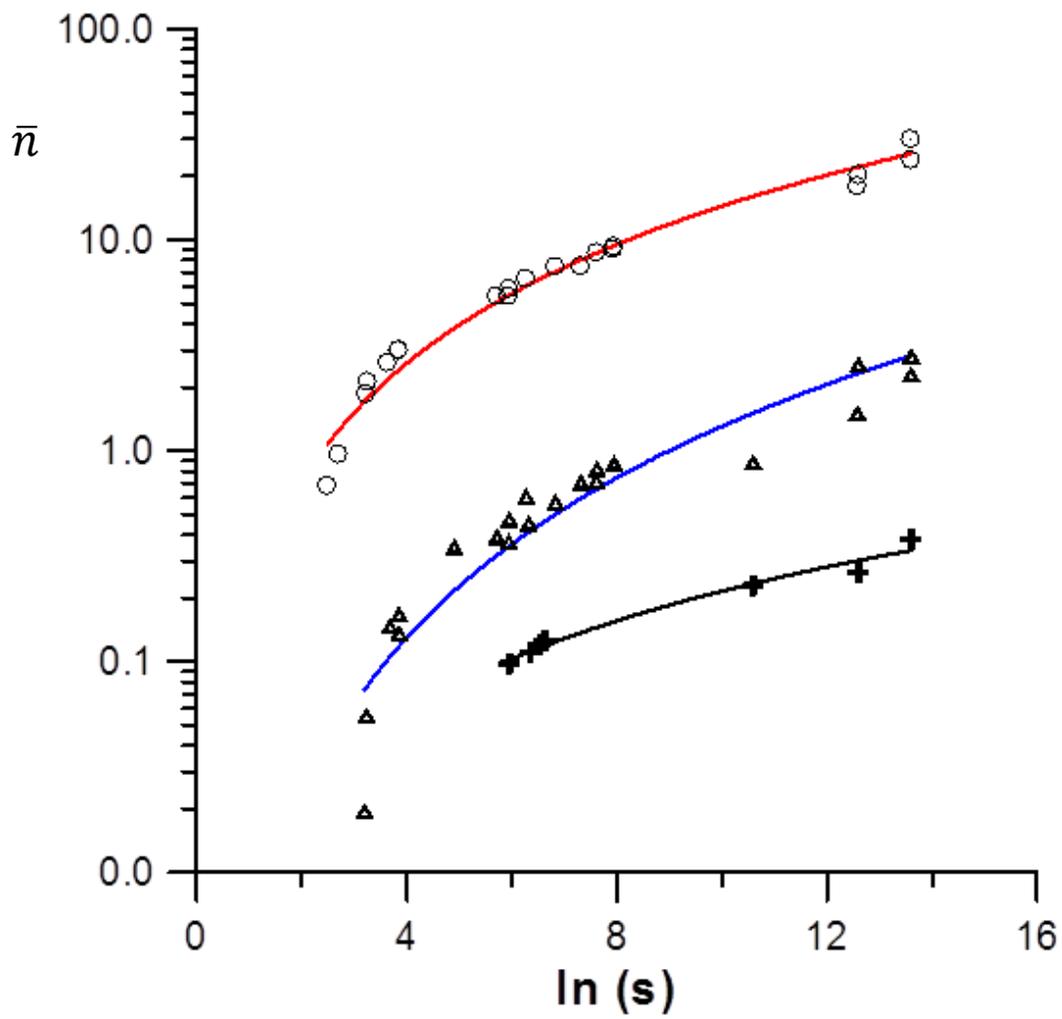

Fig (3)

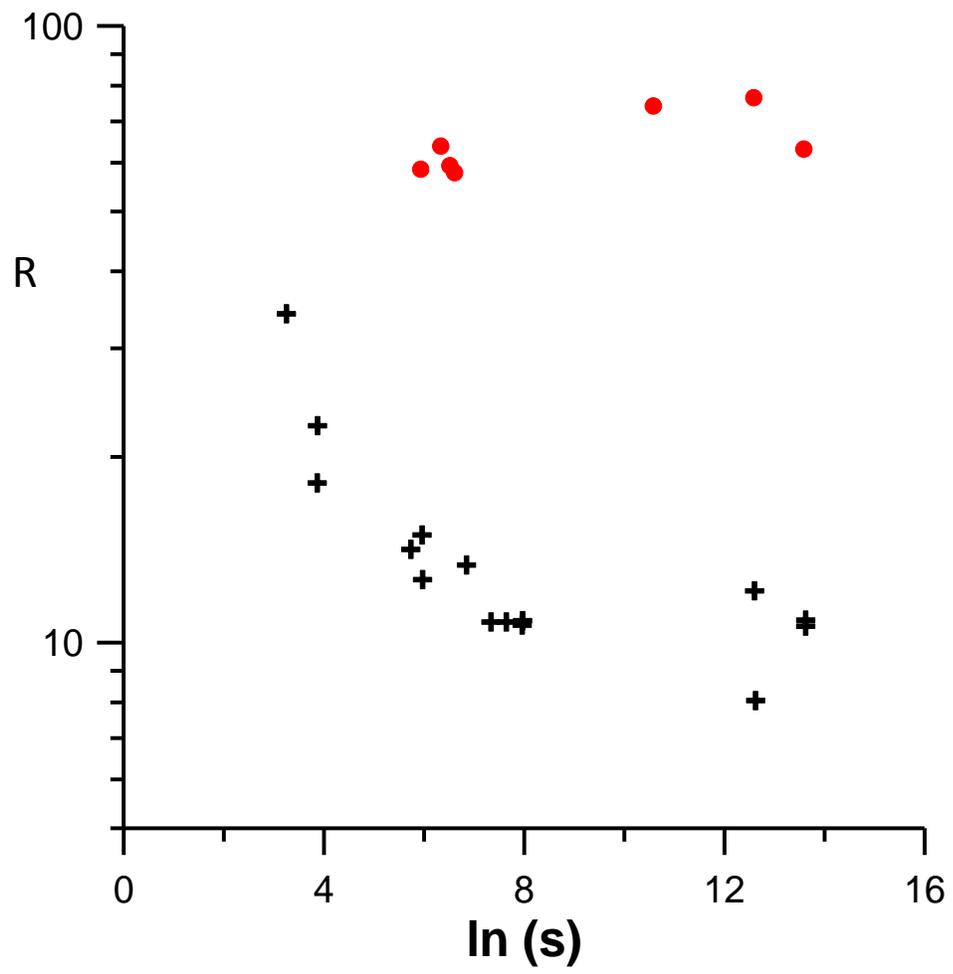

Fig (4)

# Figure Captions

**Fig.(1):** the mean multiplicity of created particles, $\bar{n}$, as a function of ln(s). Crosses are experimental data. the straight line is the fitting with the low energy part of the data, the blue curve is the fitting with whole data and the red curve represents the semi-hard and the hard classes of the events, respectively.

**Fig.(2):** Multi-pomeron exchange diagram.

**Fig.(3):** the mean multiplicities of created pions, $\bar{n}_\pi$, keons, $\bar{n}_k$ and lambdas $\bar{n}_\Lambda$ (circles, triangles and crosseses; respectively) as a function of ln(s) and the red, blue and black lines are their fittings, respectively.

**Fig.(4):** the dependence of the ratio, R, of mean multiplicities: pion to keon $R_{\pi/k}$ (black crosses), and pion to lambda $R_{\pi/\Lambda}$ (red blobs) on ln(s).